\documentclass[aps,nofootinbib,floatfix,12pt]{revtex4}
\usepackage{setspace}

\usepackage{wrapfig}
\usepackage{graphicx}

\graphicspath{{converted_graphics/}}
\begin{document}
\title{The Origin of the Universe as Revealed Through the Polarization of the Cosmic Microwave Background}         % Enter your title between curly braces
%\author
\begin{abstract}\begin{singlespace}{\tiny Scott Dodelson, Richard Easther, Shaul Hanany, Liam McAllister,
Stephan Meyer, Lyman Page, 
Peter Ade,
Alexandre Amblard,
Amjad Ashoorioon,
Carlo Baccigalupi,
Amedeo Balbi,
James Bartlett,
Nicola Bartolo,
Daniel Baumann,
Maria Beltran,
Dominic Benford,
Mark Birkinshaw,
Jamie Bock,
Dick Bond,
Julian Borrill,
François Bouchet,
Michael Bridges,
Emory Bunn,
Erminia Calabrese,
Christopher Cantalupo,
Ana Caramete,
Carmelita Carbone,
Sean Carroll,
Suchetana Chatterjee,
Xingang Chen,
Sarah Church,
David Chuss,
Carlo Contaldi,
Asantha Cooray,
Paolo Creminelli,
Sudeep Das,
Francesco De Bernardis,
Paolo de Bernardis,
Jacques Delabrouille,
F.-Xavier Désert,
Mark Devlin,
Clive Dickinson,
Simon Dicker,
Michael DiPirro,
Matt Dobbs,
Olivier Dore,
Jessie Dotson,
Joanna Dunkley,
Cora Dvorkin,
Hans Kristian Eriksen,
Maria Cristina Falvella,
Dave Finley,
Douglas Finkbeiner,
Dale Fixsen,
Raphael Flauger,
Pablo Fosalba,
Joseph Fowler,
Silvia Galli,
Evalyn Gates,
Walter Gear,
Yannick Giraud-Heraud,
Krzysztof Gorski,
Brian Greene,
Alessandro Gruppuso,
Josh Gundersen,
Mark Halpern,
Jean-Christophe Hamilton,
Masashi Hazumi,
Carlos Hernandez-Monteagudo,
Mark Hertzberg,
Gary Hinshaw,
Christopher Hirata,
Eric Hivon,
Richard Holman,
Warren Holmes,
Wayne Hu,
Johannes Hubmayr,
Kevin Huffenberger,
Howard Hui,
Lam Hui,
Kent Irwin,
Mark Jackson,
Andrew Jaffe,
Bradley Johnson,
Dean Johnson,
William Jones,
Shamit Kachru,
Kenji Kadota,
Jean Kaplan,
Manoj Kaplinghat,
Brian Keating,
Reijo Keskitalo,
Justin Khoury,
Will Kinney,
Theodore Kisner,
Lloyd Knox,
Hideo Kodama,
Alan Kogut,
Eiichiro Komatsu,
Arthur Kosowsky,
John Kovac,
Lawrence Krauss,
Hannu Kurki-Suonio,
Jean-Michel Lamarre,
Susana Landau,
Charles Lawrence,
Samuel Leach,
Louis Leblond,
Adrian Lee,
Erik Leitch,
Rodrigo Leonardi,
Julien Lesgourgues,
Andrew Liddle,
Eugene Lim,
Michele Limon,
Marilena Loverde,
Philip Lubin,
Enrico Lunghi,
Joseph Lykken,
Carolyn MacTavish,
Antonio Magalhaes,
Davide Maino,
Victoria Martin,
Sabino Matarrese,
John Mather,
Harsh Mathur,
Tomotake Matsumura,
Pieter Meerburg,
Alessandro Melchiorri,
Laura Mersini-Houghton,
Amber Miller,
Michael Milligan,
Kavilan Moodley,
Michael Neimack,
Hogan Nguyen,
Alberto Nicolis,
Ian O'Dwyer,
Angela Olinto,
Luca Pagano,
Enrico Pajer,
Bruce Partridge,
Timothy Pearson,
Hiranya Peiris,
Marco Peloso,
Francesco Piacentini,
Michel Piat,
Lucio Piccirillo,
Elena Pierpaoli,
Davide Pietrobon,
Giampaolo Pisano,
Levon Pogosian,
Dmitri Pogosyan,
Nicolas Ponthieu,
Lucia Popa,
Clement Pryke,
Christoph Raeth,
Subharthi Ray,
Christian Reichardt,
Sara Ricciardi,
Paul Richards,
Antonio Riotto,
Graca Rocha,
John Ruhl,
Benjamin Rusholme,
Robert Scherrer,
Claudia Scoccola,
Douglas Scott,
Carolyn Sealfon,
Emiliano Sefusatti,
Neelima Sehgal,
Michael Seiffert,
Leonardo Senatore,
Paolo Serra,
Sarah Shandera,
Meir Shimon,
Peter Shirron,
Jonathan Sievers,
Joe Silk,
Kris Sigurdson,
Robert Silverberg,
Eva Silverstein,
Suzanne Staggs,
Glenn Starkman,
Albert Stebbins,
Federico Stivoli,
Radek Stompor,
Naoshi Sugiyama,
Daniel Swetz,
Andrea Tartari,
Max Tegmark,
Peter Timbie,
Maxim Titov,
Matthieu Tristram,
Mark Trodden,
Gregory Tucker,
Jon Urrestilla,
Marcella Veneziani,
Licia Verde,
Joaquin Vieira,
Terry Walker,
David Wands,
Scott Watson,
Steven Weinberg,
Rainer Weiss,
Benjamin Wandelt,
Bruce Winstein,
Edward Wollack,
Mark Wyman,
Amit Yadav,
Ki Won Yoon,
Olivier Zahn,
Matias Zaldarriaga,
Michael Zemcov,
Jonathan Zwart\\
}{\it This white paper was assembled by Scott Dodelson with input from many of the cosigners. It
is part of the efforts of NASA'a Primordial Polarization Program Definition Team (PPPDT), Shaul Hanany chair, and of a NASA award to Steve Meyer and colleagues entitled "A study for a CMB Probe of Inflation"
(07-ASMCS07-0012). 
}\end{singlespace}
\end{abstract}        % Enter your name between curly braces
\date{\today}          % Enter your date or \today between curly braces
\maketitle
\thispagestyle{empty}

\newpage

\setcounter{page}{1}\pagestyle{plain}

\section*{\large Executive Summary}

Modern cosmology has sharpened questions posed for millennia about the origin of our cosmic habitat. The age-old questions have been transformed into two pressing issues primed for attack in the coming decade:

\begin{itemize}
  \item {\bf How did the Universe begin?}\\
  The current cosmological paradigm successfully explains how the majestic structure observed in the Universe today grew out of small ripples in the density of matter. What is the physical origin of the primordial seeds which are ultimately responsible for the existence of galaxies, stars, planets, and people in the Universe? It is natural to expect (and many theories predict) that whatever produced the density ripples also produced gravity waves -- undulations in the fabric of space-time which travel at the speed of light. Does the Universe contain a spectrum of primordial gravity waves produced by the same mechanism which produced the ripples in the density?

  \item {\bf What physical laws govern the Universe at the highest energies?}\\
  All explanations for the seeds of structure rely on physics at energies far beyond those probed by, e.g., CERN's Large Hadron Collider. Experiments probing these seeds therefore may provide information about new particles, forces, or perhaps even extra dimensions of space that are visible only at the highest energies.
%%
%is that in its earliest moments, the Universe underwent a tremendous burst of expansion, known as inflation. What is the underlying physical cause of inflation and how is it  related to the other laws of nature? The physical processes responsible for inflation -- or even more revolutionary alternatives -- probably involved energies a trillion times larger than will be probed at  Cosmological observations thus provide a unique probe of physics at the highest imaginable energies.
%  \item {\bf If the revolutionary idea of inflation is correct, what is the nature of the early dark energy responsible %for this epoch of accelerated expansion?} \\%
%  All of our current knowledge is consistent with the idea that 
%   \item Probing Inflation
  \end{itemize}  

The clearest window onto these questions is the pattern of polarization in the Cosmic Microwave Background (CMB), which is uniquely sensitive to primordial gravity waves. A detection of the special pattern produced by gravity waves would be not only an unprecedented discovery, but also a direct probe of physics at the earliest observable instants of our Universe. Experiments which map CMB polarization over the coming decade will lead us on our first steps towards answering these age-old questions. 

\section{How Did the Universe Begin?}

Over the course of billions of years, perturbations in the early Universe were amplified by gravitational instability, transforming  an almost perfectly smooth Universe into one with planets, stars, galaxies, and galaxy clusters. This cosmic evolution has been quantitatively confirmed: the small initial perturbations encoded in the CMB have just the right amplitude to produce the structure observed in the Universe today. We are %beginning to understand the origin of these primordial seeds of structure, and we are %emboldened to seek an understanding of the origin of the Universe itself.
emboldened to seek an understanding not only of the origin of the primordial perturbations which seeded structure in the Universe, but ultimately of the origin of the Universe itself.

Beyond their amplitude, the initial perturbations present several distinctive features~\cite{BaumannInflation}. They are nearly {\it scale-invariant}: perturbations at all wavelengths have nearly the same amplitude. They are almost exactly {\it Gaussian\/}, in that their statistical properties conform to a classic Gaussian random field to at least one part in 1000. Most strikingly, measurements of the CMB indicate that the perturbations were {\it synchronized} at early times: when the perturbations are decomposed into Fourier modes, one finds that every mode began with the same temporal phase. 

%This early synchronization is apparently imposed when perturbations were too large to fit inside the universe, defined %by the distance  light can travel since the beginning of time.

This early synchronization is particularly puzzling since it was locked in when the relevant spatial scales were apparently  larger than the distance light traveled since the beginning of time (the horizon). 
%Yet somehow these super-horizon modes were synchronized. 
This discovery of the last decade sharpens the classic horizon problem: why does radiation arriving from opposite ends of the Universe share the same temperature? The problem is now even more profound: how were the initial perturbations, with their puzzling synchronization, produced? What physical mechanism could have possibly planted these primordial seeds?

\section{New Laws of Physics}

Over the next decade, the era during which the seeds of structure were produced -- perhaps $10^{-35}$ seconds after the Big Bang -- will join nucleosynthesis (3 minutes) and recombination (380,000 years) as windows
into the primordial Universe that can be explored via present-day observations. However, recombination and nucleosynthesis depend on the well-tested details
of atomic and nuclear physics respectively, while the energy scale at which the seeds were laid down is likely to be so high that the fundamental constituents of the universe and the laws of nature at that time are currently unknown. 
%The new physics responsible for seed production likely lies at energies around one trillion times greater than those which will be studied at the Large Hadron Collider.
Our ability to see through this new window will turn the early universe into a laboratory
for ultra-high energy physics~\cite{BaumannInflation} at scales entirely inaccessible to conventional terrestrial
experimentation.

Is the new physics associated with the Grand Unified Scale at which the three low-energy forces -- weak, electromagnetic, and strong -- become one? Supersymmetry is a theory of particle physics which explains why the electroweak scale is so different from the scale associated with gravity. Is the new physics part of a supersymmetric theory? Are there other particles or fields that can be discovered which are related to those which generated the primordial perturbations? Almost all models for these seeds predict an epoch of acceleration in the early universe. Did some early form of dark energy drive this acceleration? 
%If so, is it related to the dark energy which is causing %the universe to accelerate today? 
A number of models rely on extra dimensions. Does the universe have more than three spatial dimensions? 
Almost all models rely on assumptions about the laws of physics at energies close to the Planck scale, the scale at which quantum-mechanical fluctuations render general relativity unstable.  The underlying complete theory that describes physics at the Planck scale -- perhaps a string theory, or perhaps some theory not yet conceived -- then dictates the amplitude of the gravitational waves produced.  In particular, the symmetries of this fundamental theory can leave traces in the primordial gravity wave signal, so that a detection of, or constraints on, primordial gravity waves could provide the first observational clue as to the nature of quantum gravity.

\section{Inflation}

The general considerations outlined above are most easily illustrated in the context of the most-studied model of the early Universe: inflation -- the idea that the Universe expanded nearly exponentially rapidly very early in its history. Inflation resolved several classical problems in cosmology and correctly predicted the observed features of the primordial perturbations. The early accelerated expansion drove small regions that had been in causal contact far away from one another. Quantum fluctuations, usually observed only on microscopic scales, were stretched to astronomical sizes and promoted to cosmic significance as the seeds of large scale structure. 
The wavelengths of these fluctuations became so large -- larger even than the horizon -- that the perturbations froze at a constant amplitude. When they re-entered the horizon much later, all modes were therefore synchronized to have the same temporal phase. Most models of inflation are driven by an almost constant energy density (similar to the models for dark energy today), so perturbations in the small wavelength modes which left the horizon latest were generated under the same conditions that existed when large wavelength modes left the horizon. Hence, the spectrum of perturbations is nearly scale-invariant, in agreement with observations. Additionally, the huge growth eliminated curvature, in full agreement with today's percent-level measurements that the Universe is flat.

All models of inflation make predictions for the shape of the density spectrum, the amplitude and shape of the gravity wave spectrum, and the level of deviations from Gaussianity. Many of the simplest models predict an appreciable gravity wave signal but no detectable deviations from Gaussianity, while alternatives to inflation seem to predict a Universe with no detectable primordial gravity waves but often appreciable non-Gaussianity.
%Figure \ref{fig:nsr} shows several classes of models in the plane of (spectral index of %primordial density perturbations, amplitude of gravity waves) that is ripest for% %observational attack. Indeed, Fig.~\ref{fig:nsr} shows that one of the most popular early %models (inflation driven by a scalar field with a $\phi^4$ potential) is now excluded. 
%Measurements which accurately determine these properties of the primordial perturbations will differentiate among %competing models of inflation or falsify the theory in favor of alternatives. 
The amplitude of primordial gravity waves therefore provides a way to distinguish between simple models of inflation and alternative proposals for the dynamics of the early Universe.

%\begin{figure}[tbp] % float placement: (h)ere, page (t)op, page (b)ottom, other (p)age
%  \centering
  % file name: C:/Documents and Settings/dodelson/My Documents/CMB/nsr.eps
%  \includegraphics[width=4.67in,keepaspectratio]{nsr}
%  \caption{\footnotesize Inflationary predictions (red curves and points) for, and WMAP %constraints (black curves) on, spectral index of the scalar perturbations and the amplitude% 
%of tensor perturbations (gravity waves) compared to scalar perturbations.}
%  \label{fig:nsr}
%\end{figure}

%Therefore, the gravity wave amplitude is the most powerful probe of the physics driving inflation. 
%Many models predict a large amplitude that can be detected in the coming decade. 
Moreover, the gravity wave amplitude is directly tied to the energy scale during inflation, so a detection 
%or constraint on this quantity 
can be translated into clues about the new physics responsible for the origin of structure in the Universe. The amplitude of the gravity wave spectrum is expressed relative to that of the density perturbation spectrum by the parameter $r$. Current experiments constrain $r<0.3$, and in the coming decade values of $r$ at least as low as 0.01 will be attainable. 
This amplitude of gravity waves represents a crucial target: theoretical models with $r>0.01$ are qualitatively different from those with small $r$. Particle physicists have recently made progress understanding the symmetries underlying these two classes of theories~\cite{BaumannInflation}, so detection of or constraints on $r$ will provide information about the underlying principles governing the physics operating at ultra-high energies. 
%If gravity waves are not detected at this level, then the search for nongaussianities -- predicted by many low %gravity-wave models -- will take on added importance. These two approaches are therefore very complementary.

Summarizing the reasons why the hunt for primordial gravity waves is so compelling, a detection would:
\begin{itemize}
  \item Rule out alternatives to inflation,
  \item Pinpoint the energy scale at which inflation took place,
  \item Provide clues about the symmetries underlying new physics at the highest energies.
  \end{itemize}

%so-called {\it large-field models} -- those in which the field driving inflation changes considerably in Planck units -- predict values of $r$ above this, while small-field models predict lower values of $r$. Particle physicists have recently made progress understanding the symmetries of the underlying theories that might lead to these two classes of models~\cite{BaumannInflation}, so there is the real hope that detection or constraints on $r$ will provide information about the underlying principles governing the physics operating at ultra-high energies. 
%For example, the information would constrain string theory, a proposed unification of gravity and quantum mechanics, %since only some constructions contain the relevant shift symmetry.

\section{CMB Polarization: The Ultimate Gravity Wave Detector}

%** 
Primordial gravity waves leave a unique imprint on the microwave background
\begin{wrapfigure}{r}{65mm}
  \begin{center}
    \includegraphics[width=65mm]{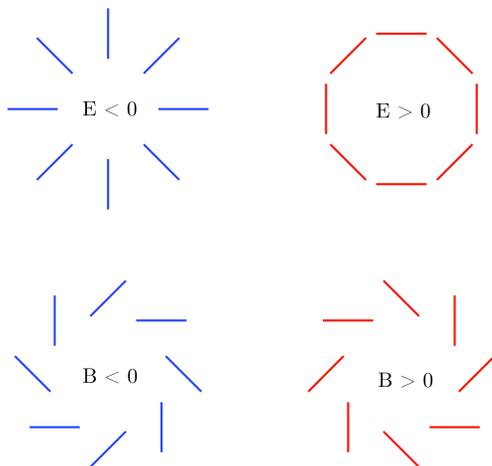}
  \end{center}\label{fig:eb}
  \caption{\footnotesize Any polarization field can be decomposed into two modes. Positive (negative) E-modes surround hot (cold) spots. B-modes cannot be produced by ordinary perturbations to the density but are produced by gravity waves.}
\end{wrapfigure}
 as they stretch and squeeze the space in which the electrons and photons interact. A quadrupole intensity anisotropy in the radiation field produces observable polarization in the CMB via Compton scattering. 
When gravity waves are the source of the anisotropy, the ensuing polarization pattern
has a handedness,  depicted as the {\it B-modes} in Figure~1.  
On the other hand, density perturbations sourcing the anisotropy produce only {\it E-mode\/} polarization patterns.   On large angular scales, the most plausible cosmological sources of a  B-mode signal are primordial gravity waves, so the amplitude of the B-mode signal is a direct measure of the gravity wave background, and thus the energy scale of inflation. A detection would be not only an unprecedented discovery, but also a direct probe of physics at the earliest observable instants of our Universe.

\begin{figure}[tb] % float placement: (h)ere, page (t)op, page (b)ottom, other (p)age
  \centering
  % file name: C:/Documents and Settings/dodelson/My Documents/CMB/EPICV2.eps
  \includegraphics[width=4.in,keepaspectratio]{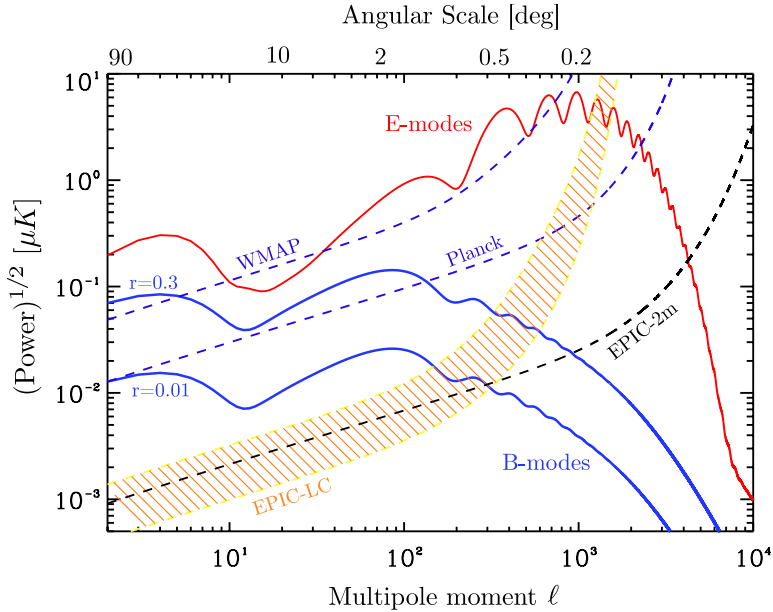}
  \caption{\footnotesize Predicted spectra of E- and B-modes. The blue solid curves representing B-modes labeled r=0.3 and r=0.01 correspond to amplitudes just below current limits and within reach of a satellite mission dedicated to polarization, respectively. The hatched region and the dashed curve labeled ``EPIC'' show the noise levels projected for two possible implementations of this mission~\cite{epic}. The dashed curves labeled ``WMAP'' and ``Planck'' correspond to the statistical noise limits
for these satellites after 9 years and 1 year, respectively. All noise curves are averaged over bins of width $\Delta l=0.3l$.}
  \label{fig:EPICV2}
\end{figure}

Figure~\ref{fig:EPICV2} depicts the expected %**
angular %** 
spectra of the two modes of CMB polarization. E-modes have been detected and a number of experiments are on the verge of pinning down their spectrum, thereby further constraining cosmological parameters. The  primordial B-mode spectrum has a characteristic double-humped shape, the first bump on large angular scales produced at the end of the Dark Ages and the second on degree scales produced during electron-photon decoupling around the time of recombination. The amplitude of the B-mode spectrum is unknown since inflationary models make a range of predictions for the amplitude of the primordial gravity waves. 
%** 
There are no known technical limitations~\cite{weiss} to achieving the sensitivity necessary to detect $r$ down to $10^{-3}$. Astrophysical foregrounds will likely degrade this sensitivity, but a variety of simulations using multiple techniques shows that a robust detection of $r$ down to a level of 0.01 -- a key threshold delineating the theoretical models -- is achievable with a future satellite mission~\cite{DunkleyFGs}. 

Beyond this principal science, CMB polarization measurements will also impact upon non-inflationary science.  These measurements will determine the gravitational potential along the line of sight to the 
last scattering surface~\cite{SmithLensing}, thereby constraining models of dark energy and possibly detecting the decaying gravitational potentials produced by massive neutrinos. CMB polarization will also constrain reionization, which heralds the end of the Dark Ages~\cite{ZaldarriagaReionization}, and will provide information about the distribution of magnetic fields in and outside our Galaxy~\cite{FraisseFGs}.

\section{Conclusion}

Cosmic microwave background polarization offers an extraordinary opportunity to gain a first glimpse into the physics that shaped our Universe. Experimentalists have demonstrated that a coordinated attack on this problem over the coming decade will likely detect primordial gravity waves -- thereby providing extensive information about new physics at ultra-high energy scales -- or severely constrain the scenario responsible for the origin of the Universe.

% Set the ending of a LaTeX document

\begin{thebibliography}{1}

\bibitem{BaumannInflation}
D.~Baumann {\it et al.}  [CMBPol Study Team Collaboration],
  ``CMBPol Mission Concept Study: Probing Inflation with CMB Polarization,''
  arXiv:0811.3919 [astro-ph].


\bibitem{weiss}
  J.~Bock {\it et al.},
  ``Task Force on Cosmic Microwave Background Research,''
  \newblock (2006), astro-ph/0604101.

\bibitem{DunkleyFGs}
 J.~Dunkley {\it et al.},
  ``CMBPol Mission Concept Study: Prospects for polarized foreground removal,''
  arXiv:0811.3915 [astro-ph].


\bibitem{epic}
J.~Bock {\it et al.},
  ``The Experimental Probe of Inflationary Cosmology (EPIC): A Mission Concept
  Study for NASA's Einstein Inflation Probe,''
  arXiv:0805.4207 [astro-ph].



\bibitem{SmithLensing}
K.~M.~Smith {\it et al.},
  ``CMBPol Mission Concept Study: Gravitational Lensing,''
  arXiv:0811.3916 [astro-ph].


\bibitem{ZaldarriagaReionization}
M.~Zaldarriaga {\it et al.},
  ``CMBPol Mission Concept Study: Reionization Science with the Cosmic
  Microwave Background,''
  arXiv:0811.3918 [astro-ph].


\bibitem{FraisseFGs}
A.~A.~Fraisse {\it et al.},
  ``CMBPol Mission Concept Study: Foreground Science Knowledge and Prospects,''
  arXiv:0811.3920 [astro-ph].



\end{thebibliography}
\end{document}